\documentstyle[prl,aps,multicol,graphics]{revtex} 

\newcommand{\figref}[1]{Fig.\ \ref{#1}}
\newcommand{\Dblk}{\Delta_{o}}

\title{Energetics and geometry of excitations in random systems}
\author{A. Alan Middleton}
\address{Department of Physics, Syracuse University, Syracuse, New York 13244}
\date{July 24, 2000}
\draft
\begin{document}

\maketitle

\begin{abstract}
Methods for studying droplets in models with
quenched disorder are critically examined.
Low energy excitations in two dimensional models are investigated
by finding minimal energy interior excitations and by computing the
effect of bulk perturbations.
The numerical data support the assumptions of compact droplets and a single
exponent for droplet energy scaling.
Analytic calculations show how strong corrections to power laws can
result when samples and droplets are averaged over.
Such corrections can explain apparent discrepancies
in several previous numerical results for spin glasses.
\end{abstract}
\pacs{75.10.Nr, 74.60.Ge, 02.70.Lq, 02.60.Pn}

\begin{multicols}{2}
Magnets and superconductors are examples of
physical systems where quenched disorder
often plays a dominant role. Such systems can exhibit hysteresis
effects and long relaxation times that are the manifestation of
the large energy barriers created by the quenched disorder.
One scenario that makes predictions for
the equilibrium and nonequilibrium
behavior of disorder dominated phases
is the droplet or scaling picture \cite{FisherHuseBrayMooreMcMillan}.
Predictions in this scenario follow from scaling assumptions for the
energetic and geometric properties of excitations.
For simple topological reasons,
excitations can be defined as regions where
the configuration is uniformly related by a symmetry to a
global ground state (e.g., spin flipped domains.)
In the droplet picture,
the low lying excitations of size $l$ are connected and compact:
they have volume $\sim l^{d_f}$, with dimension
$d_f$ equal to the system dimension $d$,
and the surface to volume
ratio decreases as $l$ increases.
Droplet boundaries are fractal, with surface dimension $d_s<d$.
The central ansatz is that the
probability distribution $\rho(\Delta,l)$
for the energy $\Delta$ of a droplet of size $l$
in a given volume $\sim l^d$ has a 
characteristic scale $\sim l^\theta$.
This distribution is argued to have finite weight at
$\Delta = 0$.
The two exponents $\theta$ and $d_s$ can be used, for example,
to predict many of
the properties of a spin glass \cite{FisherHuseBrayMooreMcMillan}.
This scenario is consistent with numeric results for
excitations created by modifying
boundary conditions \cite{SG2,McNamaraMiddletonZeng}.
However,
other work \cite{Kawashima,Riegeretal,HoudayerMartin} has
suggested that there may be more than one important energy scaling exponent
and more complicated geometries for excitations.
The proposed distinct exponents separately describe
(i) boundary induced domain wall excitations and (ii)
excitations induced by
internal constraints or external fields. It has also
been suggested that there is distinct scaling for large droplets created
by modifying the quenched disorder \cite{PalassiniYoungdroplet}.
It is important to understand these claims, as they suggest that
the standard droplet picture is, at best, incomplete.

To provide perspective, it is useful to investigate
in detail systems which lend themselves to precise study,
where some analytic results are known and large systems can be
simulated efficiently.
Results are presented here for 
a 2D elastic medium and a 2D Ising spin glass.
Single interior droplets, which include
a specified central point, are computed for the elastic medium.
In contrast with work on interior droplets in
2D spin glasses \cite{Kawashima},
a fast, exact algorithm is used,
allowing for precise checks of scaling.
The responses of the elastic medium and the spin glass
to bulk perturbations are also calculated.
The numerical results for droplet energies and geometrical characteristics
show that logarithmic or small power law corrections are quite strong.
These corrections can be understood in detail
by arguments within the droplet picture.
Droplets that are not induced by boundary conditions
are only bounded above by the system size $L$
and below by a discretization scale, so that {\em all scales
between must be considered when computing averages.}
Corrections to scaling for droplets of fixed scale
$l$, such as $l^{-1}$ or $L^{-1}$ corrections (e.g., from
lattice discreteness) or unknown irrelevant operators, might also
be considered.
However, the scale averaging corrections are
apparently dominant for some quantities.
Such corrections lead to an
effective energy exponent distinct from $\theta$,
as boundary condition induced domain walls
do not have such corrections.
To remove scale averaging corrections, one can
{\em group the droplets by scale $l$}
and study the geometry and energy as a function of $l$
(or $l/L$ if one is interested in large
droplets \cite{HoudayerMartin}),
as $L\rightarrow\infty$.
With this analysis, the numerical results provide
strong evidence that the droplets are ``compact'',
with fractal domain walls, and that there is a single energy exponent $\theta$.

One model that I study here is for a two dimensional
elastic medium, with scalar displacement field $u(x)$,
interacting with quenched periodic
disorder.
The continuum energy functional is
${\cal H}[u(x)] = \int d^2x\,[\nabla u(x)]^2+V(u(x),x)$,
where $V$ has short range correlations in its second argument and is
periodic in its first argument, $V(u(x)+1,x)=V(u(x),x)$.
This model has been used for vortex lattices in
superconductors, incommensurate charge density waves, and crystal growth
on a disordered substrate \cite{RG}.
The continuum model can be discretized on
a scale $a$, where the disorder
and elastic energies balance \cite{IMLOLR}.
As an effective degree of freeedom
$i$ is pinned to a preferred configuration (up to periodic shifts),
the displacements $u_i$ are of the form $n_i+\beta_i$, for integer $n_i$ and
fixed $\left\{\beta_i\right\}$.
Elastic interactions tend to minimize nearest neighbor differences in $u_i$,
with excitations of the medium being regions displaced
relative to the ground state.
Since the $u_i$ are discretized,
domain walls separate
regions relatively shifted by unit amounts.
Numerical work for zero temperature ($T=0$) has
determined properties of the ground state and
the scaling of boundary induced domain wall energies
\cite{ZengMiddletonShapir,MiddletonDefect,RiegerBlasum,ZengLeathFisher}.
Finite temperature simulations \cite{ZengLeathHwa}, both Monte Carlo
and combinatorial, have shown that the
$T=0$ phase is stable at finite $T$.
This model is thus
a useful prototype for models with finite $T$
transitions, such as the 3D spin glass \cite{3DSG}.

Another model treated here is the
2D Ising spin glass, with
Hamiltonian ${\cal H} = -\sum_{\left<ij\right>}J_{ij} s_i s_j$,
with spins $s_i = \pm 1$ on a triangular lattice and 
Gaussian distributed $J_{ij}$.
The ground states $\left\{s^0_i\right\}$ for samples in
this model were found by a combinatorial method for a standard
graph representation \cite{MiddletonDefect,barahona}.

\begin{figure}
\begin{center}\resizebox{0.9\linewidth}{!}{\includegraphics{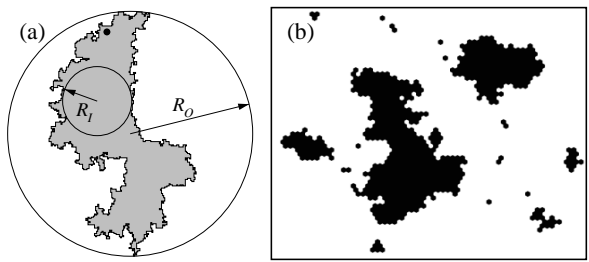}}\end{center}
\caption{
(a) Diagram of the geometry of a sample droplet in the 2D elastic medium.
The droplet is the region which
can be displaced with
minimal energy cost and
contains the sample center (dot).
The radii $R_O$ and $R_I$ are defined in the text.
(b) Droplets induced by a bulk
perturbation ($\epsilon = 16$) in
a $64^2$ spin glass sample. The filled areas have spins flipped relative
to the unperturbed ground state ($s_i^o \ne s_i^\epsilon$).
For the results shown in
Fig. \ref{sgforce}, $\epsilon\approx 0.35,1.3$ and droplets rarely intersect.
}
\label{figgeometry}
\end{figure}

For the elastic medium, minimal energy domain walls about the center of a
sample were studied on a square
lattice using a polynomial time algorithm \cite{AAMprep}
that calculated the energy $\Dblk$ and the droplet boundary.
One method to characterize the compactness of droplets is
to compare $R_O$, the radius
of the smallest circle that
encloses the droplet,
with $R_I$, the radius of the largest circle contained by the
boundary vertices.
Droplets can be studied in spin glasses by finding the ground state
and then recomputing the ground state $\left\{s_i^\epsilon\right\}$
with modified couplings
$J_{ij}\rightarrow J_{ij} - \epsilon L^{-d} s^0_is^0_j$
\cite{DrosselBokilBrayMoore,PalassiniYoungdroplet}.
This bulk perturbation can introduce excitations on all scales.
Sample excitations are depicted in \figref{figgeometry}.

The droplet energies are of great interest,
as these are believed
to determine the static correlation functions at finite temperature
and the relaxation to equilibrium.
Consider the problem of finding
the minimal energy droplet around the origin in a system
of size $L$ \cite{Kawashima}.
Assume that there is a factor $b$ which gives a separation of
length scales: droplets differing in size by $b$ are independent.
One independent droplet excitation could be excited at each
scale, so that
sums over all scales must be performed
(for similar sums,
see Refs.\ \cite{FisherBragg,DrosselBokilBrayMoore,HoudayerMartin}.)
At each length scale $s$, $s=1,2,\ldots,\log_b(L)$,
the distribution for the energy $\Delta(b^s)$
has characteristic scale $b^{s\theta}$ and has
finite weight $b^{-s\theta}$ at $\Delta = 0$
\cite{FisherHuseBrayMooreMcMillan}.
The total density of states $\rho(\Dblk)$ for $\Dblk < L^{\theta}$
is then a sum over $s$, giving
$\rho(\Dblk)\sim L^{-\theta}\left|1-(b'L)^{\theta}\right|$,
for $\theta \ne 0$,
where $b'>0$ is set by $b$ and the lattice and boundary conditions,
which
affect the lower and upper ends of the sum.
The expected minimum value for $\Dblk$ scales
as $\sim L^{\theta}\left|1-(b'L)^{\theta}\right|^{-1}$, the
subdominant term reflecting that the
minimal energy droplet is chosen from all length scales from $1$ to $L$.
The effective energy exponent is then (for $\theta < 0$)
\begin{equation}
\theta^{\rm eff}
= \frac{d\ln(\overline{\Dblk})}{d\ln(L)}
= \frac{\theta}{1-(b'L)^{\theta}}.
\end{equation}
Applying Eqn.\ (1) to the 2D spin glass,
taking $b'=2$ and $L=16$, gives an effective exponent
$\theta^{\rm eff} = -0.45$, apparently quite different from
the domain wall value $\theta=-0.28$ \cite{SG2}
and consistent with the alternate energy exponent
proposed in earlier numerical work \cite{Kawashima}.
The effective exponent
converges to $\theta$
quite slowly with $L$ (and is relatively insensitive
to $b'$), as $\theta$ is near zero.

\begin{figure}
\begin{center}\resizebox{0.9\linewidth}{!}{\includegraphics{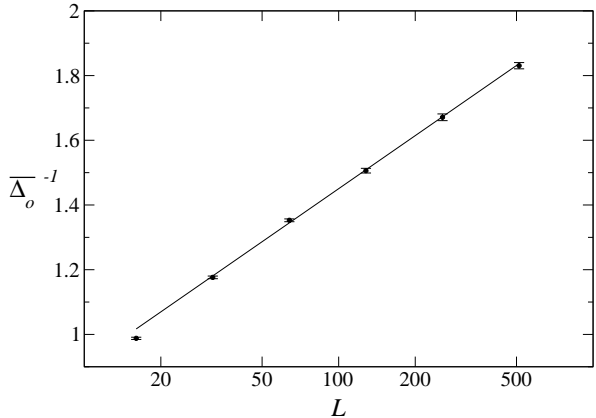}}\end{center}
\caption{
Plot of the inverse of the mean droplet energy $\overline{\Dblk}$ vs.
system size $L$ in the 2D elastic medium, averaged over
at least $10^4$ samples for each $L$. $1\sigma$ error bars are shown.
The line indicates the fit $\Dblk(L) = [0.363+0.263\,\ln(L)]^{-1}$
for $512 \ge L\ge 32$, with
$\chi^2=1.3$ for 3 d.o.f.
}
\label{figquant}
\end{figure}

One case where $\theta = 0$ for domain
walls created by boundary conditions is the 2D elastic medium.
Large domain walls can be created by external strains.
By statistical tilt invariance of the disorder \cite{statsymm},
the change in the
sample averaged energy can be found by computing the elastic energy only,
as the change in the sample averaged pinning energy is zero.
Displacing one end of a sample by $\delta u = 1$ to induce one domain wall
gives an elastic energy density $\sim L^{-2}$ over the volume $L^2$, so that
the total domain wall energy scales as a constant ($\theta = 0$.)
This result is consistent with
previous numerical simulations of boundary induced domain
walls \cite{ZengMiddletonShapir}.
However,
the mean {\em interior} droplet energy $\overline{\Dblk}(L)$ can
be fit over a decade with $-0.15 < \theta < -0.23$,
to within a few percent for smaller $L$.
Arguments similar to those for $\theta \ne 0$ can be applied to
explain this.
There are $\ln(L/a)/\ln(b)$ independent scales to choose from,
each with identical droplet energy distributions ($\theta = 0$).
The inverse of the minimal droplet energy $[\overline{\Dblk}(L)]^{-1}$
is therefore linear in $\ln(L)$.
This result can also be derived using elasticity theory.
The displacement at the origin of a region of size $a$
costs an elastic energy that scales as $\sim [\ln (L/a)]^{-1}$.
By tilt symmetry, the pinning can be averaged over, so that
$\overline{\Dblk}(L) \propto [\ln (L/a)]^{-1}$
for interior droplets constrained to contain the origin.
The numerical results
are quite consistent with these expectations, as
shown by the two parameter fit displayed in \figref{figquant}
(in addition, the computed probability of generating
a droplet of size $l$ is consistent
with a distribution uniform in $\ln(l)$ \cite{AAMprep}.)

Similar corrections are important
for the magnetization $m(h)$ of a spin
glass in response to an external field $h$ \cite{FisherHuseBrayMooreMcMillan}.
For the case $\theta \ne 0$ and
$h<O(L^{\theta-d/2})$,
$m$ is found by summing over scales
the product of the probability $hl^{d/2-\theta}$
of generating a droplet, its expected
contribution $l^{d/2}L^{-d}$ to the magnetization,
and the number of droplets $(L/l)^d$ at scale $l$.
This gives
$m\propto h\left|(b'_h L)^{-\theta} - 1\right|$,
with $b'_h > 0$ a constant,
to be compared with the uncorrected singular piece
$m\propto hL^{-\theta}$
(for numerics, see Ref.\ \cite{Riegeretal}).
The size of the corrections
are quite similar to those for $\theta$.

\begin{figure}
\begin{center}\resizebox{0.9\linewidth}{!}{\includegraphics{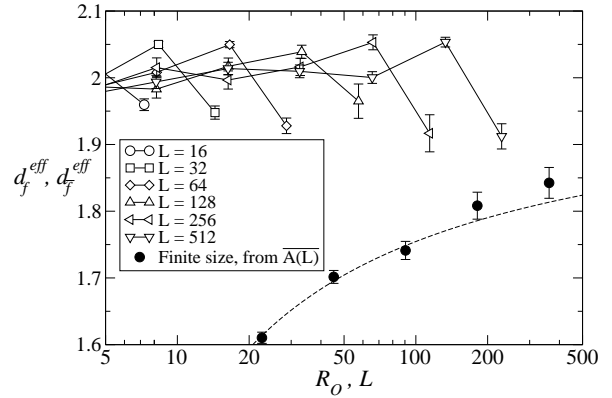}}\end{center}
\caption{
Estimates of the droplet fractal dimension
in a 2D elastic medium ($1\sigma$ statistical uncertainties shown.)
Fixing $L$, binning droplets by linear size $R_O$,
and computing the discrete logarithmic
derivative to estimate $d[\ln(A))]/d[\ln(R_O)]$
gives the
droplet dimension $d_f^{\rm eff}(R_O)$ values connected by the solid lines.
From this method $d_f=2.01(2)$.
Averaging droplet area $A$ over samples
and computing the discrete logarithmic derivative
to estimate $d[\ln(\overline{A})]/d[ln(L)]$ gives the
finite size estimate $d_{\overline{f}}^{\rm eff}(L)$. The dashed line shows
a fit using the correction derived in the text with $a'=1.7(3)$.
}
\label{figdistrib}
\end{figure}

The measurement of geometrical quantities, such as boundary length and
droplet area, can also be strongly influenced by scale averaging corrections
{\em if one averages over all length scales from $1$ to $L$.}
When the sample averaged area $\overline{A(L)}$ of interior droplets
in the 2D elastic medium
is computed as a function of $L$ and the local exponent
$d_{\overline{f}}^{\rm eff} = d[\ln(\overline{A(L)})]/d[\ln(L)]$
is computed, the local dimension is less than two,
which might suggest fractal droplets.
This local exponent slowly changes
with $L$, though (Fig.\ \ref{figdistrib}.)
The local exponent for $\overline{A(L)}$ is
$d_{\overline{f}}^{\rm eff} = 2 - [\ln(L/a')]^{-1}+O(L^{-2})$,
where $a'$ depends on the boundary and lattice cutoffs.
A useful procedure to reduce the corrections to $d_f^{\rm eff}$
is to {\em separate out the
scales} and plot the droplet area $A(R_{O},L)$
as a (binned) function of $R_{0}$.
Changing the order of the averages gives
local exponents that are much better fit by a constant.
For $L > 32$ and $R_{0} > 8$,
$\overline{A(R_o)} \sim R_o^{d_f}$, with bulk droplet dimension
$d_f = 2.01(2)$ (Fig.\ [\ref{figdistrib}].)
A similar plot for the perimeter
confirms \cite{MiddletonDefect} that $d_s = 1.25(1)$,
with the surface to volume ratio vanishing as $l^{d-d_s}$ for large droplets.
Droplet compactness can also be confirmed by plotting the ratio
$k=R_O/R_I$ binned according to $R_O$; it is found that the distribution
of $k$ converges, with
mean $\left<k(R_O,L)\right>=2.85(5)$, when $L/2 > R_O > 32$.

\begin{figure}
\begin{center}\resizebox{0.9\linewidth}{!}{\includegraphics{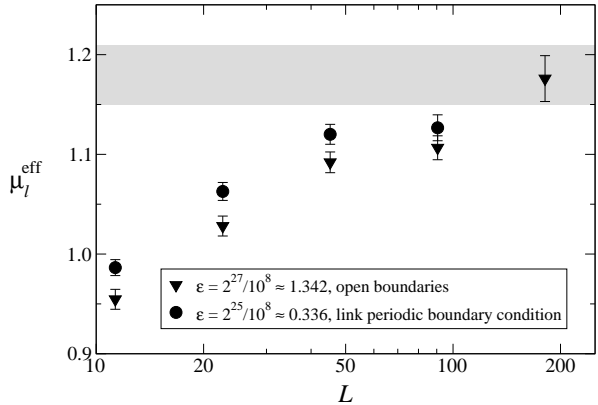}}\end{center}
\caption{
Plot of the local exponent
$\mu_l^{\rm eff}$ (from the discrete derivative
between $L=8,16,\ldots,256$)
for the link overlap $q_l$ in the 2D spin glass
as a function of system size. Results are shown for distinct $\epsilon$
and boundary conditions (open, with boundary spins fixed under perturbation,
and link periodic.)
The expected
limit at large $L$ is $\mu_l = 1.18(3)$ (shaded region.)
}
\label{sgforce}
\end{figure}

Corrections to geometric and energetic
quantities are also important when computing
link overlaps, such as those found in comparing the unperturbed $J_{ij}$
ground state, $\left\{s_i^0\right\}$, with the $\epsilon$-perturbed state.
The link overlap $q_l$ is the fraction of link
values $s_is_j$ on bonds $\left<ij\right>$ which are unchanged.
By summing the contributions over all scales (note the small droplets
in Fig.\ \ref{figgeometry}(a)), it can be shown that the local exponent
for the fraction $1-q_l$ of changed bonds behaves as
\begin{equation}
\mu_l^{\rm eff} = -\frac{d[\ln\overline{(1-q_l)}]}{d\ln(L)}
= \mu_l - \frac{c'}{(cL)^{d-\mu_l}-1},
\end{equation}
where $\mu_l=\theta+2(d-d_s)$
and $c,c'$ are constants characterizing the
upper and lower cutoffs.
The computed local exponent for the 2D spin glass
is shown in Fig.\ \ref{sgforce} ($\mu^{\rm eff}_l$
is relatively insensitive to $\epsilon$,
at least for $0.35 < \epsilon < 1.3$.)
Only for $L > 100$ does $\mu_l^{\rm eff}$ approach the large $L$ limit
of $\mu_l \approx 1.18$ (using the values $\theta=-0.28$ and
$d_s=1.27$.)
Similar results are found for the 2D elastic medium \cite{AAMprep}.
The exponent $\mu$ for spin overlaps $q=L^{-d}\sum s_i^os^\epsilon_i$,
with $1-q \sim L^{-\mu}$, has
smaller corrections of this form and may well be dominated
by corrections to scaling from unknown operators or inverse lengths.
Numerics \cite{AAMprep} show that $\mu$ converges much more quickly
than $\mu_l$ in the 2D spin glass.
The measurement of $\mu_l$ (relative to $\mu$) has been used by
Palassini and Young \cite{PalassiniYoungdroplet}
to conclude that a second energy exponent $\theta'$
affects the response to bulk perturbations in the 3D spin glass.
In three dimensions, the scale averaging corrections decrease more
quickly with $L$, as
$d -\mu_l \approx 1.3$ compared with the 2D
correction exponent $d-\mu_l \approx 0.82$,
but the system sizes that can be simulated are
much smaller.
It may be that corrections due to irrelevant variables
or possible $1/L$ effects are dominant,
but scale averaging corrections clearly {\em contribute} to errors in $\mu_l$.
A correction of $\delta\mu_l \approx -0.2$ for $L\approx 8$,
from similar $c$ and $c'$,
would invalidate the
conclusions of Ref.\ \cite{PalassiniYoungdroplet}.

In summary, 
analysis of numerical data provides a precise confirmation of the droplet
picture in the bulk of a sample, both for the scaling of the energies and
for the geometrical structure of droplets. In comparing the
numerical results with the droplet picture, care must be taken
to understand where strong corrections might arise.
The corrections arising from averaging over
multiple scales can be predicted in some detail
and apply to
spin glasses and other models where $\theta$ is near zero.
The corrections to averaged quantities such
as the droplet energy $\overline{\Dblk}(L)$
are satisfactorily explained, for $L^2>10^2$ systems,
by averaging simple power laws over scales between $1$ and $L$.
The link overlap, which inherently averages over scales, is
strongly affected by finite size corrections for
small $\theta$ and $d-d_s$, with
effective exponent corrections greater than
$0.1$ for $L < 30$ in the 2D spin glass.
It has been suggested that, for topological reasons,
distinct $\theta$ exponents exist only in $d \ge 3$ \cite{HoudayerMartin}.
The numerics in this Letter are for $d=2$,
but they and the general analysis suggest
that large finite size effects strongly affect results in three dimensions.
To reduce these types of corrections, data can be binned over
droplet size $l$ (or over $l/L$) at fixed system size $L$,
checking for convergence by then increasing $L$.

I would like to thank David Huse for
bringing to my attention some recent work on droplets,
Olivier Martin for a stimulating discussion, and Daniel Fisher for
discussions.
The LEDA library \cite{LEDA} was most useful in the study of the
geometry of the droplets.
This work was supported by the National Science
Foundation (DMR-9702242).

\end{multicols}
\end{document}